\documentclass[journal,onecolumn]{IEEEtran}
\usepackage{url}
\usepackage{float} 
\usepackage{amsfonts,amsthm,amsmath}
\usepackage{multirow}
\usepackage[linesnumbered,ruled]{algorithm2e}
\usepackage{lipsum,mathtools}
\usepackage{todonotes}
\newcommand{\FF}{\mathbb{F}}

\newtheorem{lem}{Lemma}
\newtheorem{prop}{Proposition}

\newtheorem{examp}{Example}
\newtheorem{rem}{Remark}

\usepackage{hyperref}
\hypersetup{
  colorlinks   = true,    
  urlcolor     = blue,    
  linkcolor    = blue,    
  citecolor    = red      
}

\begin{document}

\title{
{\small This work has been submitted to the IEEE for possible publication. \\
Copyright may be transferred without notice, 
after which this version may no longer be accessible.\\}
On decoding of a specific type of self-dual codes}


\author{\IEEEauthorblockN{Radinka Yorgova}\\
\IEEEauthorblockA{\textit{
 TU Delft, The Netherlands} \\
radinka.yorgova@gmail.com
}
}

\maketitle

\begin{abstract}
This work introduces a decoding strategy for binary self-dual codes possessing an automorphism of a specific type. The proposed algorithm is a hard decision iterative decoding scheme. The enclosed experiments show that the new decoding concept performs error correction beyond the upper bound for the code correction capability. Moreover, we prove that the requirements for the new algorithm hold for any binary self-dual code having an automorphism of the specific type, which makes decoding of this large group of codes possible.
\end{abstract}


\section{Introduction}

\IEEEPARstart{T}{he} existing general decoding schemes as the nearest neighbor decoding or the syndrome decoding are only applicable for short-length self-dual codes \cite[p.41]{book_codes}\footnote{In the optimal case, for a binary $[n,k]$ code both decoding schemes could use a lookup table with $2^{n-k}$ elements in $\FF_2^n$.}.
When increasing the size of the code, the required memory for each of these algorithms limits their practical applications. 
For several particular self-dual codes, there exist decoding schemes. For example, decoding by hand the $[24,12,8]$ binary extended Golay code by Pless~\cite{VPless_decoding} or decoding binary extremal self-dual code of length 40 by Kim and Lee~\cite{Kim_decoding}. Moreover, Gaborit et al. presented decoding schemes for few binary doubly-even self-dual codes of length 32~\cite{Gaborit_decoding}. 

There is no efficient general decoding algorithm for self-dual codes or for a large family of such codes to the best of our knowledge. This was defined by Pless and Huffman in 2003 as a research problem~\cite[Research Problem 9.7.8]{book_codes}.

Here, we propose a decoding algorithm for a large group of binary self-dual codes, namely self-dual codes having an automorphism of odd order. The algorithm is a hard decision iterative decoding scheme and can be used for any code of this group regardless of its length and minimum distance. 

We first provide preliminaries with definitions and results used further in this work. Then the idea in our decoding scheme is introduced. Further, we prove that the proposed algorithm is a valid algorithm for any self-dual code possessing an automorphism of odd order. Next, three examples of extremal and optimal self-dual codes are presented together with the simulation results for our decoding scheme. Two of these codes are known, whereas the third one is new and constructed to illustrate that the codes, which can benefit from the new decoding scheme, are a large group. 
At last, we discuss some open questions, possible applications, and limitations of the proposed decoding scheme.

\section{Preliminaries}

In this section, we present first the terminology related to self-dual codes in general and afterward related to self-dual codes with automorphisms of odd order.

\subsection{Self-dual Codes}

Let $\FF_2^n$ be the standard $n$-dimensional vector space over the binary field $\FF_2$ and $\cal C$ be a $k$-dimensional subspace of  $\FF_2^n$. Then $\cal C$ is a binary $[n,k]$ code. The \textit{Hamming weight} $wt(v)$ of a vector $v \in \FF_2^n$ is the number of the nonzero coordinates of $v$. If $d$ is the minimum weight of all the nonzero codewords of $\cal C$, then $\cal C$ is called a binary $[n,k,d]$ code and $d$ - minimum weight of the code. \textit{Hamming distance} between two vectors in $\FF_2^n$ is the number of coordinates where they differ.

The inner product in  $\FF_2^n$ is given by 
$u\cdot v = u_1v_1+u_2v_2+\dots+u_n v_n$ for $u,~v \in {\FF_2}^n$.
Two vectors are orthogonal if their inner product equals to $0$. Then, the orthogonal of the code $\cal C$ is ${\cal C}^\perp = \{v \in \FF_2^n ~|~ u\cdot v = 0 , ~~\forall u \in \cal C  \}$.
 
If $\cal C \subset {\cal C}^\perp$, the code $\cal C$ is called \textit{self-orthogonal} and if $\cal C= {\cal C}^\perp$, $\cal C$ is called \textit{self-dual}. It is know that the weight of any codeword of a self-dual code is even. 
A binary self-dual code $\mathcal{C}$ is called {\em doubly-even} if the weight of every codeword is divisible by four, and 
{\em singly-even} if there is at least one codeword of even weight not divisible by $4$, i.e., weight $\equiv 2 \pmod 4$ \cite[p.11]{book_codes}. 

Upper bounds for the minimum weight of a self-dual
$[n,n/2,d]$ code are given in~\cite{Rains}:
\begin{equation}
\label{bound3}
d \leq 4 \lfloor{ \frac{n}{24}}\rfloor + 4,~~ if~~n\not\equiv 22 ~(\bmod~ 24),
\end{equation}
 and
\begin{equation}
\label{bound4}
d\leq 4 \lfloor \frac{n}{24} \rfloor + 6,~~ if~~n\equiv 22 ~(\bmod~ 24).
\end{equation}


Stricter bounds are known for some specific lengths, like for $n=78$ the maximum $d$ is $14$ instead of $16$~\cite{d_bounds}.
Self-dual codes which reach the minimum weight bounds are called  \textit{extremal} whereas 
self-dual codes with the highest minimum weight for a given length among the known such are called \textit{optimal}.
In~\cite{Zhang}, it is proven that binary extremal double-even self-dual codes do not exist for lengths $n>3\,928$.

Two binary codes are equivalent if one can be obtained from the other by a permutation of coordinates. That is, if there exist a permutation $\sigma$, $\sigma \in S_n$ ($S_n$- the symmetric group of degree $n$), such that $\sigma(\cal C) = \cal C'$. If $\sigma(\cal C) = \cal C$ for some $\sigma \in S_n$, then the permutation $\sigma$ is an automorphism of the code $\cal C$. 

\subsection{Self-dual Codes and Automorphisms of Odd Order} 
\label{s_d_codes_with_autom}

Following  the notations in~\cite{Yorgova_2008}, a permutation of order $\mathcal{L}$, having $f$
fixed points and $t_1$ cycles of length $a_1$, $t_2$ cycles of length $a_2$, $\dots$, $t_h$ cycles of length $a_h$, with
$1<a_1<a_2< \dots < a_h$, is called a \textit{permutation of type $\mathcal{L}$-$(t_1,t_2,\dots,t_h;~ f)$}.
 
In this work, we only consider permutations of type $\mathcal{L}$-$(t_1;~ 0)$, where $\mathcal{L} = pr$, for $p$ and $r$ being odd primes. 
That is, $\sigma$ is of order $pr$ and has only cycles of length $pr$ and no fixed points. Thus, without loss of generality $\sigma$ can be represented as:
\vspace{-0.1cm}
\begin{equation}\label{sigma_pr}
\begin{array}{ll}
  \sigma  =&\,\Omega_{1}\Omega_{2}\ldots\Omega_{t_1},
\end{array}
\end{equation}\vspace{-0.4cm}
\\
where $\Omega_s$ is a cycle of length $pr$ for $1\leq s \leq t_1$.

Let further $\cal C$ be a binary self-dual $[n, n/2, d]$ code with a generator matrix $G$ and $\sigma$ be an automorphism of $\cal C$.

If $v\in \cal C$, then $v$ can be presented as 
$$v = ( v\vert\Omega_{1},v\vert\Omega_{2},\dots, v\vert\Omega_{t_1}),$$
where 
$v\vert\Omega_{i}=(v_0,v_1,\dots,v_{pr-1})$ denotes the coordinates of $v$ in the $i^{-th}$ cycle of $\sigma$. 
Then, the image of $v$, $\sigma(v)$, is a vector obtained from $v$ by cyclic shift in each of $v\vert\Omega_{i}$. 
From another side $\sigma(v) \in \cal C$ since $\sigma$ is an automorphism of $\cal C$. Thus,  all vectors obtained from $v$ by cyclic shifts of the coordinates  in each  $\Omega_{i}$ are also codewords.
Therefore, a matrix in the form 
\begin{equation}
\label{G_cells}
\left (
\begin{array}{llll}
 G_1& G_2 & \dots & G_{t_1}\cr
\end{array}
\right ),
\end{equation}
where $G_j$ is a circulant matrix of length $pr$, generates codewords and then it can be considered  as a sub-matrix of $G$.

Further, we follow the notations in~\cite{Yorgova_2008} adjusted to out particular type of automorphism (Eq.~\eqref{sigma_pr}). The sets $F_{\sigma}( \mathcal{C})$ and $E_{\sigma}( \mathcal{C})$ are defined as:

\begin{equation}
\label{F-phi}
 F_{\sigma} ( \mathcal{C}) = \{ v \in  \mathcal{C} \vert ~~  v \sigma =v \}
\end{equation}
and
\begin{equation} 
\label{E-phi}
\begin{array}{l}
E_{\sigma}(\mathcal{C})= \{  v \in \mathcal{C}\vert ~~wt(v\vert \Omega_i)\equiv 0\; (mod \; 2),\; 1\leq i\leq t_1 \},
\end{array}
\end{equation}
where $v\vert \Omega _i $ is the restriction of $v$ on $\Omega _i $.

It is known that $F_{\sigma}( \mathcal{C})$ and $F_{\sigma}( \mathcal{C})$ are subcodes of $\cal C$. Moreover, ${\cal C}=F_{\sigma}({\cal C}) \oplus E_{\sigma}({\cal C})$, where $\oplus$ stands for the direct sum of linear subspaces~\cite{Yorgova_2004}. Then a generator matrix of our self-dual code  $\cal C$ can be decomposed as: 
\begin{equation}
\label{gen_g}
 \begin{array}{ll}
 G = & \left (
   \begin{array}{l}
   X \cr
   Y 
  \end{array}  \right), 
\end{array}
\end{equation}
where $X$ is a generator matrices of $F_{\sigma}({\cal C})$ and $Y$ is a generator matrix of $E_{\sigma}({\cal C})$.

If a map $\pi$ for our particular type of automorphism is defined as 
\begin{equation}
\label{pi_pr}
 \pi:~ F_\sigma (\mathcal{C}) \to  \FF_2^{t_1},     ~~~~~ \pi (v \vert \Omega _i)=v_j ,
\end{equation}
for some  $j \in \Omega_i,$ $i=1,\dots,t_1$, then the image $\pi(F_{\sigma}({\cal C}))$ is a binary $[t_1,t_1/2]$  self-dual code \cite{Yorgova_2004}.

Let $\mathcal{P}$ denote the set of even-weight polynomials 
 in $\mathcal{R} = \FF_2[x]/ (x^{pr} -1)$  
and map $\varphi$ be the following:
\begin{equation*}
\label{varphiE}
\varphi : \; E_\phi(\mathcal{C}) ~~ \to ~~ \mathcal{P}^{t_1}, 
\end{equation*}
where $v\vert\Omega_{i}=(v_0,v_1,\dots,v_{pr-1})$ is identified with the polynomial
 $\varphi (v\vert \Omega_{i})(x)=v_0+v_1x+\cdots
+v_{pr-1}x^{pr-1}$ in $\mathcal{P}$ for $1\leq i\leq t_1$.

An inner product in $\mathcal{P}^{t_1}$ is defined as:
\begin{equation} 
\label{product P}
\langle g,h \rangle = g_1(x) h_1(x^{-1})+\dots+g_{t_1}(x) h_{t_1}(x^{-1})
\end{equation}
for every  $g, h \in \mathcal{P}^{t_1}$.

The image $\varphi (E_\sigma(\mathcal{C}))$ is a self-orthogonal code \cite[Lemma 1]{Yorgova_2008}, i.e., 
\begin{equation}\label{orthog_E}
u_1(x)v_1(x^{-1})+ \dots+u_{t_1}(x)v_{t_1}(x^{-1}) = 0,
\end{equation}\vspace{-0.2cm}\\
for  $\forall u, v \in \varphi (E_\sigma(\mathcal{C}))$.

This orthogonality and the cyclic structure of $G_i$ in Eq.~\eqref{G_cells} are used in the decoding strategy described next.

\section{Hard-decision Iterative Decoding Algorithm}

Let $\cal C$ and $\sigma$ be defined as before.
Let also a codeword $c\in \cal C$ be transmitted and $r = c + e$ be received where $e$ is the error vector.
In polynomial representation this is:
$(r_1(x),r_2(x),\dots,r_{t_1}(x))$ = $(c_1(x),c_2(x),\dots,c_{t_1}(x))$ $ +$ $(e_1(x),e_2(x),\dots,e_{t_1}(x))$, where $r_i(x)$, $c_i(x)$, $e_i(x)$ are in $\mathcal{R} = \FF_2[x] / (x^{pr} -1)$.

We denote by $w$ the inner product of the received $r$ with any minimum weight codeword $b \in \cal C$:
$$w(x) = \langle r,b \rangle = \sum\limits _{i=1}^{t_{1}}{r_{i}(x) {b_{i}(x^{-1})}}  ~~~ mod(x^{pr}-1)$$

Since  $r = c + e$ it follows that:
$$w(x) =\langle r,b \rangle = \sum\limits _{i=1}^{t_{1}}{c_{i}(x) {b_{i}(x^{-1})}} + \sum\limits _{i=1}^{t_{1}}{e_{i}(x) {b_{i}(x^{-1})}}.$$

The term $\sum\limits _{i=1}^{t_{1}}{c_{i}(x) {b_{i}(x^{-1})}}$ is equal to zero if $c, ~b \in  \varphi(E_\sigma(\cal C))$. In Section~\ref{proof}, we prove that this term is zero for any two codewords. Here, we only formulate the statement.

\begin{lem} 
\label{lemma1}
Let $\cal C$ be a binary $[n,n/2,d]$ self-dual code possessing an automorphism $\sigma$ of type $pr$-$(t_1;~0)$,  where $p$ and $r$ are odd primes and $n = pr t_1$.
Then, Eq.~\eqref{orthog_E} holds for every $u, ~v \in \cal C$.
\end{lem}


Thus, in the expression for $w$ only the last term remains: 

$w(x) =\langle r,b \rangle = \sum\limits _{i=1}^{t_{1}}{e_{i}(x) {b_{i}(x^{-1})}}  ~~~ mod(x^{pr}-1).$\\

It is clear that if the error vector $r$ is zero, i.e., $e_i(x)=0$ for $1\leq i\leq t_1$, then $w$ is also zero.

Let the support of $b_1(x)$, $supp(b_1(x))$,  be the set $\{\beta_1,\beta_2,\dots,\beta_d\}$ which means 
that $b_1(x) = x^{\beta_1}+x^{\beta_2}+\cdots +x^{\beta_d}$. Then $w$ can be written as:
\begin{equation*} \label{eque_w_2}
\begin{array}{rl}
w(x) = &\sum\limits _{i=1}^{t_{1}}{e_{i}(x) b_{i}(x^{-1})}  ~~~ \bmod(x^{pr}-1) \\
& \\
=&  e_{1}(x) b_{1}(x^{-1})+\sum\limits _{i=2}^{t_{1}}{e_{i}(x) b_{i}(x^{-1})}  ~ \bmod(x^{pr}-1) \\
& \\
 = &e_{1}(x) x^{-\beta_1}+e_{1}(x) x^{-\beta_2}+\cdots + e_{1}(x) x^{-\beta_d}+ \\
& \\
& +\sum\limits _{i=2}^{t_{1}}{e_{i}(x) b_{i}(x^{-1})} ~~~ \bmod(x^{pr}-1)\\
\end{array} 
\end{equation*}

If $e_1(x)=x^{\epsilon_1}+ x^{\epsilon_2}+\cdots+x^{\epsilon_r}$, then the expression for $w$ can be further reorganized as:

\begin{equation} 
\label{eque_w_3}
\begin{array}{lll}
w(x)& = &x^{\epsilon_1-\beta_1}+ x^{\epsilon_2-\beta_1}+\cdots+x^{\epsilon_r-\beta_1} +\\
& & x^{\epsilon_1-\beta_2}+ x^{\epsilon_2-\beta_2}+\cdots+x^{\epsilon_r-\beta_2} + \\
& & x^{\epsilon_1-\beta_3}+ x^{\epsilon_2-\beta_3}+\cdots+x^{\epsilon_r-\beta_3} +\\
&&  \vdots  \\ 
&&  x^{\epsilon_1 -{\beta_d}} + x^{\epsilon_2-{\beta_d}} +\cdots+ +x^{\epsilon_r-{\beta_d}}+\\  
& & +\sum\limits _{i=2}^{t_{1}}{e_{i}(x) b_{i}(x^{-1})} ~~~ \bmod(x^{pr}-1), \\

\end{array} 
\end{equation}
where all $\epsilon_i-\beta_j$ are in {\rm mod}$(pr)$.

Eq.~\eqref{eque_w_3} can be seen as: the first row contains all error positions of $e_1(x)$ shifted by $- \beta_1$ positions; the second row - all error positions of  $e_1(x)$ shifted by $- \beta_2$ positions and so on, the $d^{th}$  row- all error positions of  $e_1(x)$ shifted by $- \beta_d$ positions. The rest can be considered by the same way regarding the error positions of $e_2$, $e_3$, $\dots, e_{t_1}$ with shifts corresponding to the supports of $b_2, b_3, \dots, b_{t_1}$. Note that many of these terms can be canceled out since these shifted error positions can be the same for different $e_ib_i$.

$~~$

If we multiply $w$ by $x^{\beta_1}$, the first row of $x^{\beta_1} w(x)$ in Eq.~\eqref{eque_w_3} will become exactly $e_1(x)$. Multiplying $w$ with $x^{\beta_2}$, 
the second row in Eq.~\eqref{eque_w_3} will become exactly $e_1(x)$ and so on. 

Thus, each of the polynomials:
\begin{equation}
\label{beta_w}
x^{\beta_1}w(x),~ x^{\beta_2}w(x), ~\dots,~x^{\beta_{d}}w(x)~~ ~~{\rm mod}(x^{pr} -1) 
\end{equation}
contains $e_1(x)$,  $~x^{\beta_j - \beta_s}e_1(x)$, and $x^{\beta_j }\sum\limits _{i=2}^{t_{1}}{e_{i}(x) b_{i}(x^{-1})}$ ${\rm mod}(x^{pr}-1)$, which are  
the original and shifted error positions of $e_1(x)$ and shifted error positions of $e_i(x)$, $i=2,3,\dots,t_1$. Some of the original or shifted error positions of $e_1(x)$  can be canceled out with some of the shifted error positions of $e_2(x), e_3(x), \dots,e_{t_1}(x)$.
If in all polynomials $x^{\beta_s}w(x)$ in Eq.~\eqref{beta_w}, we count the number of 1s in each position $1,2,\dots,pr$, it is expected some of the error positions to have higher frequency than the rest of the positions. 
The same process can be repeated for 
$$e_{2}(x) b_{2}(x^{-1}), ~ e_{3}(x) b_{3}(x^{-1}), ~\dots, ~ e_{t_1}(x) b_{t_1}(x^{-1}).$$ 



In general, if we consider $e_{s}(x) b_{s}(x^{-1})$, where $supp(b_s(x)) = \{ \beta_1^{(s)}, \beta_2^{(s)},\dots,\beta_d^{(s)}\}$, then the polynomials:
\begin{equation}
\label{beta_w2}
x^{\beta_1^{(s)}}w(x), ~ x^{\beta_2^{(s)}}w(x), ~\dots,~x^{\beta_{d}^{(s)}}w(x)~~ ~~{\rm mod}(x^{pr} -1)
\end{equation}
contain error positions and shifted error positions of $e_s(x)$ and shifted error positions of 
$e_i(x)$, $i=1,2,\dots,t_1$, $i\neq s$.
Therefore, when counting the number of 1s in each position from 1 till $pr$ in the polynomials in Eq.~\eqref{beta_w2},
a higher number of 1s will be an indicator for an error in this position. 
The polynomials in  Eq.~\eqref{beta_w2} are created for every $s$, $1 \leq s \leq t_1$ and the  number of 1s in each position in included in $\Phi_j^{(s)}$ defined latter.
\\

Let $\cal C$ has $L$ cyclicly different minimum weight codewords. In our case \textit{cyclically different} codewords means that one cannot be obtained from the other by applying $\sigma^t$, for some $t$, that is, $b \neq \sigma^t(c)$ for $1\leq t \leq pr-1$, $\forall b,~c \in \cal C$. The counting of 1s in each position from 1 till $pr$ in the cycles $\Omega_j$ is repeated for the polynomials in Eq.~\eqref{beta_w2} for all cyclicly different minimum weight codewords $b$. The number is denoted by $\Phi_j^{(s)}$:
\begin{equation} 
\label{phi_j_new}
\begin{array}{ll}
   \Phi_j^{(s)}  =&\sum\limits_{l=1}^{L}~ \sum\limits_{i \in supp (b_s^{(l)}(x))}  {w^{(l)}_{i+j~\bmod pr } } ~, \\
   &\\
    &j = 0,1,2,\cdots,pr-1, ~~ s=1,2,\dots,t_1, 
\end{array}
\end{equation}
$~~$\\
where $w^{(l)}_{i+j~\bmod pr}$ is counted in position $j$ since the shift of $w(x)$ by $-i$ moves the position $i+j~\bmod pr$ into $j$.

As it was mentioned, a higher number of 1s in a position from 1 till $pr$ in the polynomials in Eq.~\eqref{beta_w2} is an indicator for an error in this position. Thus, an error is expected in  the $j^{-th}$ position in cycle $\Omega_{s}$ if $\Phi_j^{(s)}$ has a higher value than the sum for any other position in the same cycle $\Omega_{s}$ and than the sum on any position in the other cycles. 

$~~$


Note that the idea of shifting and counting 1s in each coordinate position is introduced in~\cite{bossert_cyclic}, and it is specifically and only for cyclic codes. The presence of the cyclic cells in the generator matrix of the code $\cal C$ and moreover, the orthogonality in polynomial representation of each two codewords make it possible to use shifting and counting in a similar way.

$~~$  

Since $w(x) = \langle r,b \rangle  = \sum\limits _{i=1}^{t_{1}}{r_{i}(x) b_{i}(x^{-1})}$ $~~~ mod(x^{pr}-1)$, one can determine the coefficients $w_i$ of the polynomial $w(x)$ as linear combinations of the coefficients $r_i^{(t)}$ of $r_i(x)$ and the supports of $b_i(x)$ for $1\leq i \leq t_1$. 

$~~$

Here, we define our decoding strategy. We suppose that $\cal C$ and $\sigma$ are defined as before and $G$ is a generator matrix of $\cal C$. As noted above, a high value of $\Phi_j^{(s)}$ indicates an error in position $j$ of $\Omega_s$.\\
\\
\textbf{Decoding Scheme~\footnote{A \textit{pseudocode} of the decoding scheme is provided in Appendix~\ref{appendix2}.}}\\
\\
1) generate  all or almost all cyclically different codewords of weight $d$ or $d+o$ for some small $o$, for example 2 or 4 or 6. Denote the set by $D_1$; \\ 
\\
2) for the received vector $r$ compute $r G$. If $r G=0$, then $r$ belongs to the code $\cal C$ $\rightarrow$ end, otherwise $\rightarrow$ 3);\\
\\
3) split  $r$ into $t_1$ polynomials of $\FF_2[x]/(x^{pr}-1)$, i.e., $r=(r_1(x),r_2(x),\dots,r_{t_1}(x))$; \\
\\
4) for $r=(r_1(x),r_2(x),\dots,r_{t_1}(x))$ 
compute:\\
\\
$~~~~\bullet~$ $w(x) = \langle r,b \rangle  = r_{1}(x) b_{1}(x^{-1}) + r_{2}(x) b_{2}(x^{-1}) + \cdots $ $+ r_{t_1}(x) b_{t_1}(x^{-1})$ $~ {\rm mod}(x^{pr}-1)~ $ \\
\\
$~~~~~~~~$ for $~\forall ~b\in D_1$,  $b=(b_1(x),b_2(x),\dots,b_{t_1}(x))$
\\
\\
$~~~~\bullet~$ $x^{\beta_{i}^{(1)}}w(x)$ $ ~ {\rm mod}(x^{pr}-1)~ $  for $~\forall~\beta_i^{(1)} \in supp(b_1(x))$  \\
\\
$~~~~~~~~$ $x^{\beta_{i}^{(2)}}w(x)$ $ ~ {\rm mod}(x^{pr}-1)~ $  for $~\forall ~\beta_i^{(2)} \in supp(b_2(x))$\\
$~~~~~~~~$ $\vdots$ \\
\\
$~~~~~~~~$ $x^{\beta_{i}^{(t_1)}}w(x)$ $ ~ {\rm mod}(x^{pr}-1)~ $  for $~\forall ~\beta_i^{(t_1)} \in supp(b_{t_1}(x))$
\\

Note first that, for the chosen code $\cal C$ if $r \in \cal C$, then $w(x)=0$ and second, the products $x^{\beta_{i}^{(s)}}w(x)$ are cyclic shifts of $w$ with number of positions which values are from the support of $b_s(x)$.
\\
\\
5) compute $\Phi_j^{(s)}$ defined in Eq.~\eqref{phi_j_new} for $j = 0,1,\cdots,pr-1, ~ s=1,2,\dots,t_1$;\\
\\
6) determine $\Phi_{max} = max\{\Phi_j^{(s)} ~|~ j = 0,1,\cdots,pr-1, ~ s=1,2,\dots,t_1 \}$ and find the position with the value $\Phi_{max}$, i.e., find the cycle $s_1$ and position(s) $j_1$ such that $\Phi_{j_1}^{(s_1)} = \Phi_{max}$;\\
\\
7) flip the coordinate of $r_{s_1}(x)$ by adding $x^{j_1}$ to $r_{s_1}(x)$, i.e., $r_{s_1}(x)$ becomes $r_{s_1}(x)+x^{j_1}$;\\
\\
8) for the modified $r$ repeat from 2).
\\

In this algorithm, we could use set of cyclically different codewords of weight $d$ or weight slightly higher than $d$. It is because if in the first set  the codewords have only 0 coordinates in some of the cycles of $\sigma$, it is clear that no error in this cycle can be corrected.
Also, if the first set is very small, the algorithm does not perform error correction close to the error-correction capability of the code. 
Therefore, experiments are required for finding a suitable number of low weight codewords that have good decoding performance. In Section~\ref{sec:examples} we give three examples and the decoding performance of different sets of codewords.

\section{Proof of Lemma~\ref{lemma1}} 
\label{proof}

To prove Lemma~\ref{lemma1}, we first prove that Eq.~\eqref{orthog_E} holds for every $u, ~v \in F_{\sigma}( \mathcal{C})$, then for every $u\in F_{\sigma}( \mathcal{C})$ and every $v\in E_{\sigma}( \mathcal{C})$, and at last, the statement in the lemma, for $ \forall u,v \in \cal C$.

\begin{prop}\label{prop1_new}
Let $\cal C$ be a binary self-dual code having an automorphism $\sigma$ of type $pr-(t_1;~ 0)$. Let each $u\in F_\sigma(\cal C)$ be presented as $u = (u_1(x),u_2(x),\dots,u_{t_1}(x))$. Then, for $u,~v \in F_\sigma(\cal C)$ it follows:
\begin{equation}
\label{prop1}
u_1(x)v_1(x^{-1})+ \dots+u_{t_1}(x)v_{t_1}(x^{-1}) = 0.
\end{equation} 
\end{prop}

\begin{proof}
The codewords of the subcode $F_{\sigma}({\cal C})$ can be seen as one row circulant matrices in Eq.~\eqref{G_cells}. 
Moreover, the coordinates of a fixed codeword are constant in each $pr$ cycle, i.e., $v\vert \Omega _{i} = (0,0,\dots,0)$ or  $v\vert \Omega _{i} = (1,1,\dots,1)$ for $v\in F_\sigma(\cal C)$, $1\leq i \leq t_1$. 
If a map $\pi$ is defined as in Eq.~\eqref{pi_pr}, then $\pi(F_{\sigma}({\cal C}))$ is a binary $[t_1,t_1/2]$  self-dual code. 

Therefore,  $t_1$ must be even and the weight of each element in $\pi(F_{\sigma}({\cal C}))$ must also be even. Thus, each fixed codeword in $\cal C$ has an even number of $v\vert \Omega _{i} = (1,1,\dots,1)$.
This written in polynomials in  $\FF_2[x]/(x^{pr}-1)$ is:\\
each $g = (g_1(x),g_2(x),\dots,g_{t_1}(x)) \in F_{\sigma}({\cal C})$ has an even number of coordinates $g_s(x)$ of the form $1+x+x^2+\dots +x^{pr-1}$.

If $1+x+x^2+\dots +x^{pr-1}$ is denoted by $g_0(x)$, then it is clear that 
$g_0(x^{-1}) = g_0(x)~\bmod(x^{pr}-1)$ and 
$ g_0(x)g_0(x^{-1}) = g_0(x) ~\bmod(x^{pr}-1)$.

Thus, we can conclude that Eq.~\eqref{prop1} holds for two codewords in $F_{\sigma}({\cal C})$ if they intersect in even positive number of cycle positions with full one vector or, they intersect in 0 cycle positions with full one vector. 

Let we assume that there exist two codewords $v,~ v'  \in F_{\sigma}({\cal C})$ that intersect in odd number of cycle positions with full one vector and this odd number is $z$. Since the length of the cycles is $pr$, then the regular inner product of $v$ and $v'$, 
$v.v' =  v_1v_1'+ v_2v_2'+\cdots+v_{prt_1}v_{prt_1}' ~~ \bmod (2)$, 
will be equal to $z p r$, where $z$ is odd and $p$ and $r$ are odd primes. Thus,
$v.v' = z p r  \equiv  1  ~~ {\rm mod} (2)$, which is a contradiction to $\cal C$ being a self-dual code. 
Therefore, Eq.~\eqref{orthog_E} holds for any two codewords in $F_{\sigma}({\cal C})$. 
\end{proof}

Let now $u \in F_\sigma({\cal C})$ and $v \in E_\sigma({\cal C})$, where 
$u = (u_1(x),u_2(x),\dots,u_{t_1}(x))$ with coordinates $u_i(x) = 0$ or $u_i(x) = 1+x+x^2+\dots +x^{pr-1}$, 
and $v = (v_1(x),v_2(x),\dots,$ $v_{t_1}(x))$, where $v_i(x) \in \mathcal{P}$.

Since the weight of $ \pi(u)$ is even, then $u$ contains even number of coordinates equal to $1+x+x^2+\dots +x^{pr-1}$. 
Let $u_i(x) = 1+x+x^2+\dots +x^{pr-1}$ for $i\in\{\alpha_1,\alpha_2,\dots,\alpha_{2s}\}$.
Then, $\langle u,v \rangle$ and $\langle v,u \rangle$ are:
\begin{equation*} 
\label{eque_u_v}
\begin{array}{ll}
\boldsymbol{\langle u,v \rangle}  &= \sum\limits _{i=1}^{t_{3}}{u_i(x)v_i(x^{-1})}= \cr
& \cr
 = \sum\limits _{i\in \{\alpha_1,\dots,\alpha_{2s}\}} &{(1+x+x^2+\dots +x^{pr-1})~v_i(x^{-1})}= \cr
 &\cr
 &= \sum\limits _{i\in \{\alpha_1,\alpha_2,\dots,\alpha_{2s}\}}{\frac{x^{pr}-1}{x-1} ~v_i(x^{-1})}= \cr
&\cr
&= {\frac{x^{pr}-1}{x-1} \sum\limits _{i\in \{\alpha_1,\alpha_2,\dots,\alpha_{2s}\}}v_i(x^{-1})} \cr
&\cr
\end{array}
\end{equation*}

\begin{equation*}
\begin{array}{ll}
\boldsymbol{\langle v,u \rangle} ~~~~~~~~~~~~& = \sum\limits _{i=1}^{t_{3}}{v_i(x)u_i(x^{-1})}= \cr
& \cr
 &= \sum\limits _{i=1}^{t_{3}}{v_i(x)u_i(x)}= \cr
& \cr
 &= {\frac{x^{pr}-1}{x-1} \sum\limits _{i\in \{\alpha_1,\alpha_2,\dots,\alpha_{2s}\}}v_i(x)} \cr
\end{array} 
\end{equation*}

%
%
$~~$
\\
Thus, \\
%
\begin{gather}
\nonumber \langle u,v \rangle \equiv 0~~{\rm mod}(x^{pr}-1)~ 
{\rm if ~and ~only ~if } \\
\nonumber \sum\limits _{i\in \{\alpha_1,\alpha_2,\dots,\alpha_{2s}\}}v_i(x^{-1})\equiv0 ~~{\rm mod}(x-1)
\\
{\rm and}\label{iff} \\
\nonumber\\
\nonumber \langle v,u \rangle \equiv 0 ~~ {\rm mod}(x^{pr}-1)~ {\rm if~ and~ only~ if} \\
\nonumber \sum\limits _{i\in \{\alpha_1,\alpha_2,\dots,\alpha_{2s}\}}v_i(x) \equiv 0 ~~{\rm mod}(x-1).    
\end{gather}


Let us discuss the coordinates $v_i(x)$ of $v\in E_\sigma(\cal C))$.
By definition $E_{\sigma}({\cal C})= \{ v \in{\cal C}\vert ~~{\rm wt} (v\vert\Omega_{i}) \equiv 0\,(\bmod\, 2),~~ i=1,\ldots,t_{1}\},$
which implies that the weight of $v_i(x)$ is even for $1\leq i\leq t_1$.
When $v_i(x)$ has even number of nonzero coefficients, then $v_i(1) = 0$ in $\FF_2$, which means $1$ is a root of $v_i(x)$. Therefore, $x-1$ divides $v_i(x)$ for $1\leq i\leq t_1$.

From this one can conclude that  
$$ \sum\limits _{i}v_i(x) \equiv 0 ~~{\rm mod}(x-1)$$ for any $i$ and in particular, this will also be 
satisfied for $i\in \{\alpha_1,\alpha_2,\dots,\alpha_{2s}\}$, which means that 
\begin{equation}
\label{eq1}
    \sum\limits _{i\in \{\alpha_1,\alpha_2,\dots,\alpha_{2s}\}}v_i(x) \equiv 0 ~~{\rm mod}(x-1).
\end{equation}

Is $\sum\limits _{i\in \{\alpha_1,\alpha_2,\dots,\alpha_{2s}\}}v_i(x^{-1})\equiv0 ~~{\rm mod}(x-1)$ also true?

$~~~$\\
The polynomial $v_i(x^{-1}) \equiv v_i(x^{pr-1}) \in \FF_2[x]/(x^{pr}-1)$.

$~~~$\\
Since $x\equiv 1 ~~{\rm mod}(x-1)$ it follows that
$x^{pr-1}\equiv 1 ~~{\rm mod}(x-1)$ and then,
$x^{pr-1} -1 \equiv 0 ~~{\rm mod}(x-1)$.

$~~~$\\
From above, $x-1$ divides $v_i(x)$ for $1\leq i\leq t_1$. Thus, $$v_i(x) = (x-1)v_i'(x)$$ and then, 
 $$v_i(x^{-1}) = v_i(x^{pr-1}) =$$ 
 $$= (x^{pr-1}-1)~ v_i'(x^{pr-1}) \equiv 0 ~~{\rm mod}(x-1)$$
for $1\leq i\leq t_1$.

$~~~$\\
In particular, it also holds for $i\in \{\alpha_1,\alpha_2,\dots,\alpha_{2s}\}$ and therefore 
\begin{equation}
\label{eq2}
\sum\limits _{i\in \{\alpha_1,\alpha_2,\dots,\alpha_{2s}\}}v_i(x^{-1})\equiv0 ~~{\rm mod}(x-1).    
\end{equation}

Combining Eqs. \eqref{iff},~\eqref{eq1}, and~\eqref{eq2}, we derive the following proposition.

\begin{prop}\label{prop3_new}
Let $\cal C$ be a binary self-dual code having an automorphism $\sigma$ of type $pr-(t_1;~ 0)$. Then  $u\in F_\sigma(\cal C)$,  $v\in E_\sigma(\cal C)$ are orthogonal, namely $ \langle u,v \rangle \equiv 0~~{\rm mod}(x^{pr}-1)$ and $\langle v,u \rangle \equiv 0 ~~ {\rm mod}(x^{pr}-1)$.
\end{prop}

Summarizing Proposition~\ref{prop1_new},~Proposition~\ref{prop3_new}, and~\cite[Lemma 1]{Yorgova_2008}, we can prove Lemma~\ref{lemma1}.
\begin{proof}
Let $u,~v \in \cal C$, where ${\cal C}=F_{\sigma}({\cal C}) \oplus E_{\sigma}({\cal C})$.
Then, 
$$u=u'+u'', ~~ v=v'+ v'',$$
where $u',~ v' \in F_\sigma(\cal C)$ and  $u'',~ v'' \in E_\sigma(\cal C)$.
The inner product of $u$ and $v$ is 
$$\langle u,v \rangle = \langle u'+u'' ,v'+ v'' \rangle = $$
$$ = \langle u',v' \rangle + \langle u',v'' \rangle+\langle u'',v' \rangle+\langle u'',v'' \rangle.$$
The first term $\langle u',v' \rangle=0$ according to Proposition~\ref{prop1_new}. 
The next two terms are also 0 because of Proposition~\ref{prop3_new}. The last term $\langle u'',v'' \rangle$  is also 0  because $\varphi (E_\sigma(\mathcal{C}))$ is a self-orthogonal code~\cite[Lemma 1]{Yorgova_2008}. Therefore, $\langle u,v \rangle = 0$ for any $u,~v \in \cal C$.
\end{proof}

\begin{rem}
In case the binary self-dual code $\cal C$ has an automorphism $\gamma$ of odd prime order $p$ with $c$ cycles and no fixed points (type $p-(c,0)$), the fixed subcode $F_{\gamma}(\cal C)$ is also self-dual~\cite{Huffman_82} and, the image $\varphi (E_\gamma(\mathcal{C}))$ of the even subcode $E_{\gamma}(\cal C)$ is also self-orthogonal~\cite{Yorgov_83}. Following the proofs of Proposition~\ref{prop1_new}  and Proposition~\ref{prop3_new}, one can conclude that they also hold for $p$ instead of $pr$. Therefore, Lemma~\ref{lemma1} is also a valid statement for self-dual codes with an automorphism of type $p-(c,0)$. With this, the group of the self-dual codes that can be decoded by the new decoding scheme is expanded.

\end{rem}

\section{Examples}
\label{sec:examples}

The decoding algorithm is applied on three examples of self-dual code with the required structure, where 
two of the codes are known, whereas the third one is new. The last is constructed to demonstrate that self-dual codes with automorphism of the particular type exist and it is not hard to generate them when their minimum weight is not close to its upper bound Eq.~\eqref{bound3} and Eq.~\eqref{bound4}.


\subsection{Decoding of a Binary [90,45,14]  Self-dual Code}\label{ex1}

\begin{examp} 
Let $\cal D$ be a binary $[90,45,14]$ self-dual code with an automorphism $\phi$ of type $15-(6;~0)$. Note that it is an optimal code since the upper bound for the minimum distance is $16$. The code $\cal D$ holds the conditions in Lemma~\ref{lemma1} and therefore, the decoding Algorithm~\ref{decoding_s_d} is applicable to it. The number of errors which $\cal D$ is capable to correct is  $t\leq \lfloor \frac{d-1}{2} \rfloor$ errors, i.e., maximum 6 errors.
\end{examp}

In Appendix~\ref{appendix3}, the construction of the generator matrix of the code $\cal D$, defined in Eq.~\eqref{gen_g} is included. 
\\


The code $\cal D$ has $375$ codewords of weight $14$, $11\,745$ of weight $16$ and $215\,915$ of weight $18$. The rank of these sets is $45$, $44$, and $45$, respectively. 
The cyclically different codewords are isolated and the sets are denoted by $B_{14}$, $B_{16}$, and $B_{18}$, where $|B_{14}|$ = $25$, $|B_{16}|$ = $783$ and $|B_{18}|$ =$14\,399$ (Table~\ref{table_D}).


\begin{table}[H]
\caption{Sets of codewords in the $[90,45,14]$ s-d code $\cal D$} \label{table_D}
\centering
\footnotesize
\begin{tabular}{l|l|l|l|l} 
\hline 
weight $i$ & $A_i$ & rank& cyclically different& in simulations \\ \hline
&&&& \\ [-2ex] 
14 & 375 & 45& $B_{14}$, $|B_{14}|$ = 25 & $D_1$, $|D_1| = 25$ \\[0.5ex] 
16 & 11\,745 & 44& $B_{16}$, $|B_{16}|$ = 783  & $D_2$, $|D_2| = 450$\\[0.5ex] 
18 & 215\,915 &45& $B_{18}$, $|B_{18}|$ =14\,399     & $D_3$, $|D_3| = 340$ \\[0.2ex] 
\hline 

\end{tabular}
\\$~~$
\\
\end{table}

In simulations, we use three different sets of cyclically different codewords and compare their performance.
The sets are $D_1$, $D_2$, and $D_3$, which are also given in Table~\ref{table_D}.
$~~$  \\

We perform simulations on 2\,000 random error vectors and random encoded messages
for each number of errors $t$, for $t=1,2,\dots,8$. The simulation steps are:
\begin{itemize}
  \item generate a random error vector $e$ of length $90$ and weight $t$;
  \item generate a random message vector $m$ of length $45$;
  \item encode the  message vector $m$ into $c=m G$;
  \item compute $r = e+c$;
  \item decode the received vector $r$ using the iteration steps $2$ till $8$
  of Algorithm~\ref{decoding_s_d};
\end{itemize}

$~~$

The results of the simulations are presented in Table~\ref{table_res_n90}. 
They show that the set $D_1$ of the cyclically different codewords of weight $14$ is too small for a good decoding performance.
The decoding algorithm, using each of the other two sets, $D_2$ and $D_3$,  reaches the error correcting capability of the code and corrects $100\%$ of the errors, where errors are in the range up to $6$.
 Moreover, these two sets in the experiments correct $99.95\%$ and $95.35\%$ of the cases with 7 errors, and $96.8\%$ and $60.35\%$  of the cases  with $8$ errors.
 
The results of the experiments indicate a high error-correcting capability of our new algorithm, a capability beyond the upper bound for $t$.

Note that the rank of 
the set of weight $16$ codewords is $44$ which means that it is possible in step 4) to obtain a vector $r(x)$ such that the corresponding $w(x) = 0$  but $rG \neq 0$. That is the reason we consider also the set $D_3$ that has rank $45$, and the described exception is not possible. 

\vspace{0.3cm}
\begin{table*}[ht]
\caption{Decoding performance of the $[90,45,14]$ self-dual code $\cal D$}\label{table_res_n90}
\centering
\small
\begin{tabular}
{|p{0.03\linewidth}|p{0.05\linewidth}|p{0.07\linewidth}|p{0.05\linewidth}||
p{0.03\linewidth}|p{0.05\linewidth}|p{0.07\linewidth}|p{0.05\linewidth}||
p{0.03\linewidth}|p{0.05\linewidth}|p{0.07\linewidth}|p{0.05\linewidth}|}
 \hline
\multicolumn{4}{|c||}{Decoding set $D_1$, $|D_1| = 25$} & \multicolumn{4}{c||}{Decoding set $D_2$, $|D_2| = 450$} &\multicolumn{4}{c|}{Decoding set $D_3$, $|D_3| = 340$} \\ \hline
   $t$  & $tested$ & $corrected$ & $\%$ &$t$  & $tested$ & $corrected$ & $\%$ &$t$  & $tested$ & $corrected$ & $\%$ \\
   \hline
   1& 90 & 90 & 100        & 1 & 90& 90 & 100& 1& 90 & 90 & 100 \\
   2& 2\,000 & 1\,932 & 96.6   & 2 & 2\,000 & 2\,000 & 100 &   2& 2\,000 & 2\,000 & 100  \\
   3 & 2\,000 & 1\,928 & 96.4  & 3 & 2\,000 & 2\,000 & 100  &3 & 2\,000 & 2\,000 & 100 \\
   4 & 2\,000 & 1\,955 & 97.75 & 4 & 2\,000 & 2\,000   &  100 &4 & 2\,000 & 2\,000 & 100 \\
   5 & 2\,000 & 1\,930 & 96.5  & 5 & 2\,000 & 2\,000  & 100 &5 & 2\,000 & 2\,000 & 100 \\
   6 & 2\,000 & 1\,829 & 91.45 & 6 & 2\,000 & 2\,000  & 100 &6 & 2\,000 & 2\,000 & 100 \\
   7 & 2\,000 & 1\,268 & 63.4  & 7 & 2\,000 & 1\,999  & 99.95 &7 & 2\,000 & 1\,907 & 95.35 \\
   8 & 2\,000 &  560 &  28   & 8 & 2\,000 & 1\,936 & 96.8   &8 & 2\,000 &  1\,207   & 60.35  \\
   \hline
\end{tabular}
\end{table*}


\subsection{Decoding of a Binary [78,39,14] Self-dual Code}
\label{sec_ex2}

\begin{examp} \label{ex2}
 Let $\cal T$ be a binary $[78,39,14]$ self-dual code with an automorphism $\phi_1$ of type $39-(2;~0)$. Note that $\cal T$ is an extremal code~\cite{d_bounds}.
 Since  $d=14$, the number of errors which $\cal T$ is capable to correct is up to 6. 
As in the previous example, the conditions in Lemma~\ref{lemma1} are satisfied for the code $\cal T$ and therefore, the decoding Algorithm~\ref{decoding_s_d} is applicable to $\cal T$.
\end{examp}

A generator matrix of $\cal T$ is available in Appendix~\ref{appendix3}. 

This particular example has $3\,081$ minimum weight codewords and $46\,116$ codewords of weight $16$. The rank of the sets is $39$ and $38$, respectively. Among them the cyclically different are $79$ with weight $14$ and $1\,644$ with weight $16$.

The set of $79$ elements is denoted by $T_1$ and the set of $79$ elements of weight $14$ together with $244$ elements of weight $16$ is denoted by $T_2$ ( Table~\ref{table_T}).
The sets $T_1$ and $T_2$ are used in the decoding simulations.

\begin{table}[H]
\caption{Sets of codewords in the $[78,39,14]$ self-dual code $\cal T$} \label{table_T}
\centering
\footnotesize
\begin{tabular}{l|l|l|l|l} 
\hline 
weight $i$ & $A_i$ & rank& cyclically different& in simulations \\ \hline 
&&&& \\ [-2ex] 
14 & 3\,081 & 39 & $B_{14}$, $|B_{14}|$ = 79 & $T_1$, $|T_1| = 79$ \\ [0.5ex] 
16 & 46\,116 & 38& $B_{16}$, $|B_{16}|$ = 1\,644 & $T_2'$, $|T_2'| = 244$\\[0.5ex] 
14,16 & 49\,197 &39 & $B_{14}	\cup B_{16}$     & $T_2 = T_1 \cup T_2'$ \\[0.2ex] 
\hline 

\end{tabular}
\\$~~$
\\
\end{table}

Similarly to the first decoding example, for both sets $T_1$ and $T_2$, for each $t$, $1\leq t\leq 8$, we perform simulations on 2\,000 random error vectors and random encoded messages. The simulation steps are the same. The results are included in Table~\ref{table_res_n78}.
\vspace{0.3cm}
\begin{table}[H]
\caption{Decoding performance of the $[78,39,14]$ self-dual code  $\cal T$ }\label{table_res_n78}
\centering
\footnotesize
\begin{tabular}{|c|c|c|c||c|c|c|c|}
 \hline
\multicolumn{4}{|c||}{Decoding set $T_1$, $|T_1| = 79$} & \multicolumn{4}{c|}{Decoding set $T_2$,$|T_2| = 323$}  \\ \hline
   $t$  & $tested$ & $corrected$ & $\%$ &$t$  & $tested$ & $corrected$ & $\%$ \\
   \hline
   1& 78 & 78 & 100  & 1& 78 & 78 & 100 \\
   2& 2\,000 & 2\,000 & 100&  2& 2\,000 & 2\,000 & 100 \\
   3 & 2\,000 & 2\,000 & 100  &3 & 2\,000 & 2\,000 & 100  \\
   4 & 2\,000 & 2\,000 & 100  &4 & 2\,000 & 2\,000 & 100 \\
   5 & 2\,000 & 2\,000 & 100 &5 & 2\,000 & 2\,000 & 100  \\
   6 & 2\,000 & 2\,000 & 100  &6 & 2\,000 & 2\,000 & 100  \\
   7 & 2\,000 & 1\,974  &  98.7 &7 & 2\,000 & 1\,995 &  99.75 \\
   8 & 2\,000 &  1\,530 &  76.5 &8 & 2\,000 & 1\,725 & 86.25 \\
   \hline
\end{tabular}
\end{table}

The values in Table~\ref{table_res_n78} show that the set of $14$ weight codewords are sufficient for the complete decoding of $6$ errors which is the upper bound for the error capability of this code example. Differently from the first code, here both, the set $T_1$ and $T_2$, have a high error-correcting performance beyond the upper bound for $t$.

\vspace{0.5cm}

In both decoding examples, the complete set of $16$ (or $16$ and $18$) weight cyclically different codewords is not considered. There is a trade-off between the speed and memory of the decoder from one side and the decoding performance of the algorithm from another side. To achieve  decoding up to $\frac{d-1}{2}$ errors for the first example it is sufficient to use set $D_3$ with $340$ codewords and for the second example, set $T_1$ with $79$ codewords.


\subsection{Decoding of a Binary [266,133,36] Self-dual Code}
\label{sec_ex3}

\begin{examp} \label{ex3}
Let $B$ be a binary $[266,133,36]$ self-dual code with an automorphism $\phi_2$ of type $133 -(2; 0)$. It is an optimal code since the upper bound for $d$ is $48$ (Eq.~\eqref{bound3}) and, to the best of our knowledge, there is no example of a self-dual $[266,133,d\geq 36]$.
The code $B$ has error correcting capability of $17$ errors.
Since the code possesses the specific automorphism of type $pr-(t_1;~0)$, then the conditions in Lemma~\ref{lemma1} are satisfied and the decoding Algorithm~\ref{decoding_s_d} is applicable to $B$.
\end{examp}

First, the construction of the code is presented and then the decoding experiments.

The code $B$ possesses and automorphism $\phi_2$ of type $133-(2;0)$. Then:\\
1) ${B}=F_{\phi_2}(B) \oplus E_{\phi_2}(B)$; \\
2) the fixed subcode $\pi(F_{\phi_2}(B))$ is a binary $[2,1]$ self-dual code, and \\
3)  the vectors of the image $\varphi(E_{\phi_2} (B))$ are from ${\cal P}^2$,
where ${\cal P} \subset {\FF}_2/(x^{133}-1)$.

From 2) it follows that the generator matrix of $F_{\phi_2}(B)$ is 
$X=\left (
\begin{array}{ll}
l &l \cr
\end{array}
\right ),$
where $l = (1,1,\dots,1)$  is the full one vector in $\FF_2^{133}$.

Applying 3) by computer check an example for the generator matrix $Y$ of $F_{\phi_2}(B)$ is obtained:
$$Y= 
\begin{pmatrix}
y_{1,1}  &y_{1,2} \cr
\vdots & \vdots \cr 
y_{9,1} & y_{9,2}
\end{pmatrix},
$$
where $y_{i,j}$ are right-circulant $3\times 133$ cells for the first 2 rows in $Y$ and 
$y_{i,j}$ are right-circulant $18\times 133$ cells for the next 7 rows. The first rows of these circulant matrices are given in 
Table~\ref{table_c} in Appendix~\ref{appendix3}.

For the code $B$ only part of the codewords with weight $36$, $38$ and $40$ are generated. They are denoted by $L_{36}$, $L_{38}$ and $L_{40}$, respectively (Table~\ref{table_M}).
For decoding of code $B$  we use two sets, $M_1$ and $M_2$, of cyclically different codewords.
Both of them contain elements of $L_{36}$, $L_{38}$ and $L_{40}$. Details are given in Table~\ref{table_M}.

\begin{table}[H]
\caption{Sets of codewords in the $[266,133,36]$ s-d code $B$} \label{table_M}
\centering
\footnotesize
\renewcommand{\arraystretch}{1.2}
\begin{tabular}{l|l|l} 
\hline 
weight $i$ & cyclically different  & rank \\ \hline
36 &$L_{36}$,  $|L_{36}|$ = 16  &  16 \\[0.5ex]
38 &$L_{38}$,  $|L_{38}|$ = 58  &  58 \\[0.5ex] 
40 &$L_{40}$,  $|L_{40}|$ = 2\,616  &  132 \\[0.5ex] 
36 &$M_{1,36}\subset L_{36}$,  $|M_{1,36}|$ = 13  &  13 \\[0.5ex]
38 &$M_{1,38}\subset L_{38}$, $|M_{1,38}|$ = 22  &  22 \\[0.5ex]
40 &$M_{1,40}\subset L_{40}$, $|M_{1,40}|$ = 1\,455 &  132 \\[0.5ex]
40 &$M_{2,40}\subset L_{40}$, $|M_{2,40}|$ = 2\,594 &  132 \\[0.5ex]
36,38,40 &$M_1 = M_{1,36}\cup M_{1,38}\cup M_{1,40}$ &  133 \\[0.5ex]
36,38,40 &$M_2 = L_{36}\cup L_{38}\cup M_{2,40}$ &  133 \\[0.5ex] \hline
\end{tabular}
\\$~~$
\\
\end{table}

As in the previous two examples, for the sets $M_1$ and $M_2$ and each number of errors $t$, in this case $13\leq t \leq 18$, we perform a decoding experiment on $2\,000$ received encoded messages with $t$ errors. The simulation steps described in \ref{ex1} are followed, where the length of the error vector is $266$ and the length of the message is $133$. The results of the simulations are provided in Table~\ref{table_code_B}.

\begin{table}[H]
\caption{Decoding performance of the $[266, 133, 36]$ self-dual code  $B$ }\label{table_code_B}
\centering
\footnotesize
\begin{tabular}{|c|c|c|c||c|c|c|c|}
 \hline
\multicolumn{4}{|c||}{Decoding set $M_1$, $|M_1| = 1\,490$} & \multicolumn{4}{c|}{Decoding set $M_2$,$|M_2| = 2\,614$}  \\ \hline
   $t$  & $tested$ & $corrected$ & $\%$ &$t$  & $tested$ & $corrected$ & $\%$ \\
   \hline
   13 & 2\,000 & 2\,000 & 100  &  13 & 2\,000 & 2\,000 & 100  \\
   14 & 2\,000 & 1\,945 & 97.25 &  14&  2\,000 & 2\,000 & 100 \\
   15 & 2\,000 & 1\,635 & 81.75 &  15& 2\,000 & 2\,000 & 100 \\
   16 & 2\,000 & 1\,075 & 53.75  &  16 & 2\,000 & 1\,476& 73.8  \\
   17 & 2\,000 & 543 & 27.15   &  17  & 2\,000 & 1\'040  & 52 \\
   18 & 2\,000 & 350 & 17.5   &  18 & 2\,000 &404 & 20.2  \\
   \hline
\end{tabular}
\end{table}

Code $B$ has error-correcting capability of $17$ errors. The results in  Table~\ref{table_code_B}
show that set $M_1$ is too small for decoding via the new algorithm since only $13$ errors are $100\%$ corrected whereas $17$ errors are corrected in only $27.15\%$ of the cases.
When using the set $M_2$ the new decoding Algorithm~\ref{decoding_s_d} corrects the errors upto $15$ in $100\%$ of the cases and $16$ errors in almost $74\%$. The algorithm reaches the error-correcting capability of the code $B$ in  only  $52\%$ of the cases. 
To increase the error- correction performance, a larger number of cyclically different codewords of weight $36$ and close to $36$ have to be generated and included in the decoding set.  
%
Note that  $L_{36}$, $L_{38}$ and $L_{40}$ are only part of the sets of codewords with weight $36$, $38$ and $40$. We expect that if the decoding Algorithm~\ref{decoding_s_d} uses all the cyclically different codewords of weight $36$ and $38$ together with a subset of $L_{40}$, then it will reach the error-correction capability of the code in   $100\%$ of the cases.

\section{Open Questions and Limitations} 
The proposed decoding algorithm opens some questions about its possible applications. One of those questions is if a programming implementation can reach the speed requirements for any of the current decoding applications.
Is an efficient hardware implementation possible, for example an efficient
very large scale integrated (VLSI) implementations of en/decoder? 
Another open question is if this algorithm is suitable for use in the decryption process of Code-based cryptosystems. 

Since the algorithm uses a subset of the minimum weight, or close to the minimum weight codewords, it implies a limitation for a large length extremal or optimal self-dual codes. This is because this set of codewords first has to be generated, which is computationally expensive (infeasible) for a large length and minimum distance. Moreover, extremal self-dual codes are only known for a length up to $130$. 

Regardless of the questions above, we proposed an efficient decoding algorithm for a large family of self-dual codes.

Moreover, it is well known that the automorphism groups of one of the best self-dual codes have a very large order. For example, the extended Golay code $G_{24}$ - Mathieu group $M_{24}$ of order 
$2^{10} \cdot 3^3 \cdot 5 \cdot 7 \cdot 11 \cdot 23 = 244\,823\,040$ or extended quadratic residue code $QR_{80}$-  $PSL(2,79)$ of order $2^4 \cdot 3 \cdot 5 \cdot 13 \cdot 79 = 246\,480$. This is an indicator that the new decoding scheme  would be applicable for a large set of self-dual codes with as high as possible error correcting capability.



\appendices

\section{Decoding Algorithm} 
\label{appendix2}

Algorithm~\ref{decoding_s_d} gives the pseudocode of how to decode self-dual codes having an automorphism.

\begin{algorithm}
\label{decoding_s_d}
Generate the set $D_1$ of all or almost all cyclically different codewords of weight $d$ or $d+o$ for some small $o$ (2, 4, or 6)\\
Compute $r G$. 
$~~~~~~~~~~~~~~~~~~~~~~~~~~~~~~~~~~~~~~~~~~~~~~~~~~~~~~~~~~~~~~~~~~~~~~~~~~~~~~~~~$
$~~~~~~~~~~~~~~~~~~~~~~~~~~~~~~~~~~~~~~~~~~~~~~~~~$
If $r G=0$ ($r \in\cal C$)  $\rightarrow$ end,
$~~~~~~~~~~~~~~~~~~~~~~~~~~~~~~~~~~~~~~~~~~~~~~~~~~~~~~~~~~~~~~~~~~~~~~~~~~~~~~~~~$
$~~~~~~~~~~~~~~~~~~~~~~~~~~~~~~$
 else $\rightarrow$ 3);
\\
Split  $r$ into $t_1$ polynomials of $\FF_2[x]/(x^{pr}-1)$, i.e., $r=(r_1(x),r_2(x),\dots,r_{t_1}(x))$ \\
{Compute:
\begin{itemize}
    \item  $w(x) = \langle r,b \rangle  = r_{1}(x) b_{1}(x^{-1}) + r_{2}(x) b_{2}(x^{-1}) + \cdots$\\
     $~~~~~~~~~~~~~~~~~~~$ $ + r_{t_1}(x) b_{t_1}(x^{-1}) ~ {\rm mod}(x^{pr}-1)~ $ \\ $~~$
     \\
    for $\forall ~b\in D_1$,  $b=(b_1(x),b_2(x),\dots,b_{t_1}(x))$
    \\ $~~$
    \item $x^{\beta_{i}^{(1)}}w(x)$ $ ~ {\rm mod}(x^{pr}-1)~ $ 
    for $\forall~\beta_i^{(1)} \in supp(b_1)$
    \\ $~~$
    \item $x^{\beta_{i}^{(2)}}w(x)$ $ ~ {\rm mod}(x^{pr}-1)~ $
    for  $\forall ~\beta_i^{(2)} \in supp(b_2)$
    \\ $~~$
    \item  $~~\vdots$
    \\$~~$
    \item $x^{\beta_{i}^{(t_1)}}w(x)$ $ ~ {\rm mod}(x^{pr}-1)~ $ 
    for $\forall ~\beta_i^{(t_1)} \in supp(b_{t_1})$
\end{itemize}
}
 Compute $\Phi_j^{(s)}$ defined in Eq.~\eqref{phi_j_new} for $j = 0,1,\cdots,pr-1, ~ s=1,2,\dots,t_1$;\\
Determine $\Phi_{max} = max\{\Phi_j^{(s)}\}$ and,  the cycle  $s_1$ and position(s) $j_1$ such that $\Phi_{j_1}^{(s_1)} = \Phi_{max}$\\
Compute  $r_{s_1}(x)+x^{j_1}$.$~~~~~~~~~~~~~~~~~~~~~~~~~~~~~~~$
$~~~~~~~~~~~~~~~~~~~~~~~~~~~~~~~~~~~~~~~~~~~~~~~~~~~~~~~~~~~~~~~~~~~~~~~~~~~~~~~~~$
$r_{s_1}(x)$ $\leftarrow$ $r_{s_1}(x)+x^{j_1}$\\
Repeat from 2) for the modified $r$.
\caption{Decoding self-dual codes having an automorphism}
\end{algorithm}

\section{Generator matrices for the examples}
\label{appendix3}

This section provides generator matrices for the examples as discussed in Section~\ref{sec:examples}.

\subsection{Example~\ref{ex1}} 
\label{appendix3_ex1}
Let $\cal D$ be the binary $[90,45,14]$ self-dual code of Example~\ref{ex1}.
The generator matrix of $\cal D$ defined in Eq.~\eqref{gen_g} requires the matrices $X$ and $Y$ generating the subcodes $F_{\phi}({\cal D})$ and $E_{\phi}({\cal D})$, respectively. 

For $X$ a possible choice  is
$$X=\left (
\begin{array}{llllll}
l &o &l & o &o & o \cr
o & l& o &l &o & o \cr
o& o &o & o & l & l \cr
\end{array}
\right ),$$
where $l = (1,1,\dots,1)$, $o=(0,0,\dots,0)$, i.e., the full one vector and the zero vector in $\FF_2^{15}$.

The subcode $E_{\phi}({\cal D})$ it generated via its image $\varphi( E_{\phi} ({\cal D}))$. A full description of how the subcode $\varphi( E_{\phi} ({\cal D}))$ is constructed is available in~\cite{n92_2006}. 
Here, we present one example of generator matrix of $\varphi( E_{\phi} ({\cal D}))$, namely:
$$A' = 
{\small
\left(
\begin{array}{cccccc}
e_1(x)&    0 &    0  &     0  & \mu_1(x) &  \mu_1^2(x) \\
   0  & e_1(x)&    0 & \mu_1(x)& \mu_1(x) & \mu_1^{12}(x) \\
   0  & 0     & e_1(x) & \mu_1^2(x)& \mu_1^{12}(x) & \mu_1^{8}(x) \\
e_2(x)&    0 &    0  &     0  & \mu_2(x) &  \mu_2(x) \\
   0  & e_2(x)&    0 & \mu_2(x)& \mu_2(x) & 0 \\
   0  & 0     & e_2(x)& \mu_2(x)& \mu_2(x) & \mu_2(x) \\
  0   &\mu_3(x)&\mu_3(x)&e_3(x) &     0    &   0    \\
\mu_3(x)&\mu_3(x)&\mu_3(x)&  0  & e_3(x) &     0    \\
\mu_3(x)&    0   &\mu_3(x)&  0  &   0    & e_3(x)   \\
e_4(x)&    0 &    0  &     0  & 0 &  \mu_4(x) \\
  0 &   e_4(x)&    0 &     0  & \mu_4^2(x) & 0  \\
  0 &  0      & e_4(x)& \mu_4^2(x) & 0   &   0  \\
\end{array}
\right),}$$
where $e_i$, $\mu_i$ for $1\leq i \leq 4$ are given in Table~\ref{table_a}. All the polynomials in $A'$ are even weight polynomials of $\FF_2[x]/(x^{15}-1)$.

\begin{table}[H]
\caption{Elements of $\FF_2[x]/(x^{15}-1)$}
\label{table_a}
\centering
\footnotesize
\begin{tabular}{l|l} 
\hline  & \\ [-2ex]
$e_1$ & $x^{14} + x^{13} + x^{12} +x^{11} +x^9 +x^8 +x^7 +x^6 +x^4 +$\\
        & $x^3 +x^2 +x$\\ [0.3ex]
\hline 
 & \\ [-2ex]
$e_2$ & $x^{14} + x^{13} + x^{12} + x^{11} + x^9 + x^7 + x^6 + x^3$\\ [0.3ex]
\hline  & \\ [-2ex]
$e_3$&  $x^{12} + x^9 + x^8 + x^6 + x^4 + x^3 + x^2 + x$ \\ [0.3ex]
\hline   & \\ [-2ex]
$e_4$& $x^{14} +x^{13} +x^{11} +x^{10} +x^8 +x^7 +x^5 +x^4 +x^2 +x$  \\ [0.3ex]
\hline  & \\ [-2ex]
$\mu_1$ & $x^{11} + x^{10} + x^6 + x^5 + x + 1$ \\ [0.3ex] 
\hline  & \\ [-2ex]
$\mu_2$ & $x^{14} + x^{13} + x^{11} + x^9 + x^8 + x^5 + x + 1$\\ [0.3ex] 
\hline  & \\ [-2ex]
$\mu_3$& $x^{14} + x^{10} + x^7 + x^6 + x^4 + x^2 + x + 1$\\ [0.3ex] 
\hline  & \\ [-2ex]
$\mu_4$ &  $x^{13} + x^{12} + x^{10} + x^9 + x^7 + x^6 + x^4 + x^3 + x+1$\\ [0.3ex] 
\hline 
\end{tabular}
\\$~~$
\\
\end{table}

The corresponding generator matrix of the subcode $E_{\phi}({\cal D})$ is 
$$Y= 
\begin{pmatrix}
y_{1,1} & \ldots &y_{1,6} \cr
\vdots & \ddots & \vdots \cr 
y_{12,1} & \ldots & y_{12,6}
\end{pmatrix},
$$
where $y_{i,j}$ are right-circulant $4\times 15$ cells for the first 9 rows in $Y$ and 
$y_{i,j}$ are right-circulant $2\times 15$ cells for the last 3 rows. The first rows of these circulant matrices are corresponding to the given polynomials in matrix $A'$.

\subsection{Example~\ref{ex2}} 
\label{appendix3_ex2}
Let $\cal T$ be the binary $[78,39,14]$ self-dual code with an automorphism $\phi_1$ of type $39-(2;~0)$ considered in \ref{sec_ex2}.

The the matrices $X$ and $Y$ of the generator matrix of $\cal T$ defined in Eq.~\eqref{gen_g} are:
$$X=\left (
\begin{array}{ll}
l &l \cr
\end{array}
\right ),$$
where $l = (1,1,\dots,1)$  is the full one vector in $\FF_2^{39}$ and 
$$Y= 
\begin{pmatrix}
y_{1,1} & y_{1,2} \cr
y_{2,1} & y_{2,2} \cr
y_{3,1} & y_{3,2} \cr
y_{4,1} &y_{4,2} \cr
\end{pmatrix},
$$
where $y_{i,j}$ are right-circulant $12\times 38$ cells for the first 3 rows in $Y$ and 
$y_{i,j}$ are right-circulant $2\times 39$ cells for the last row. 
The first rows of these circulant matrices are given in Table~\ref{table_b}.

\begin{table}[H]
\caption{Coefficients of polynomials in $\FF_2[x]/(x^{39}-1)$}
\label{table_b}
\centering
\footnotesize
\begin{tabular}{l|l} 
\hline
$y_{1,1}$ & 000100110100101101110101100111110111111 \\
$y_{1,2}$ & 110011010100011111000101000011010111100 \\ 
$y_{2,1}$ & 100111101011000010100011111000101011001 \\
$y_{2,2}$ & 011111101111100110101110110100101100100 \\
$y_{3,1}$ & 011010000010001101000001000110100000100 \\
$y_{3,2}$ & 011111111111101111111111110111111111111 \\
$y_{4,1}$ & 011011011011011011011011011011011011011 \\
$y_{4,2}$ & 011011011011011011011011011011011011011 \\
\hline  
\end{tabular}
\end{table}

\subsection{Example~\ref{ex3}} 
\label{appendix3_ex3}

\begin{table}[H]
\caption{Coefficients of polynomials in  $\FF_2[x]/(x^{133}-1)$}
\label{table_c}
\centering
\footnotesize
\begin{tabular}{l|l} 
\hline
$y_{1,1}$& 111010011101001110100111010011101001110100111010011 \\
        & 101001110100111010011101001110100111010011101001110\\
        & 1001110100111010011101001110100 \\ \hline 
$y_{1,2}$& 000000000000000000000000000000000000000000000000000\\
        & 000000000000000000000000000000000000000000000000000 \\
        & 0000000000000000000000000000000 \\ \hline 
$y_{2,1}$& 000000000000000000000000000000000000000000000000000\\
        & 000000000000000000000000000000000000000000000000000 \\
        & 0000000000000000000000000000000 \\ \hline 
\hline  
\end{tabular}
\end{table}

\begin{table}[H]
\centering
\footnotesize
\begin{tabular}{l|l} 
\hline
$y_{2,2}$&100101110010111001011100101110010111001011100101110\\
        & 010111001011100101110010111001011100101110010111001 \\
        & 0111001011100101110010111001011 \\ \hline 
$y_{3,1}$& 000100110101111001100111101010010011100101111111110\\
        & 110011001011100001011110101100010101111111010111000\\
        & 1111000011100100110010101000000 \\ \hline 
$y_{3,2}$&000000000000000000000000000000000000000000000000000\\
        & 000000000000000000000000000000000000000000000000000 \\
        & 0000000000000000000000000000000 \\ \hline 
$y_{4,1}$& 100111011100110111010101010100101001111010000100000\\
         & 100100001111001100100101010010110010111011010011011\\
         & 1010110010010000101110010000011\\ \hline 
$y_{4,2}$&000000010101001100100111000011110001110101111111010\\
         &100011010111101000011101001100110111111111010011100 \\
         &1001010111100110011110101100100\\ \hline 
$y_{5,1}$&011110111000111111000100111011101011010000100000111\\
         & 011001011100111011010011101010100110000010100101110\\
         & 0111110100110010111101001110100 \\ \hline 
$y_{5,2}$&100111011100110111010101010100101001111010000100000\\
        & 100100001111001100100101010010110010111011010011011 \\
        & 1010110010010000101110010000011 \\ \hline 
$y_{6,1}$&111000001001110100001001001101011101100101101110100\\
         &110100101010010011001111000010010000010000101111001\\
         &0100101010101011101100111011100\\ \hline 
$y_{6,2}$&000101110010111101001100101111100111010010100000110\\
        &010101011100101101110011101001101110000010000101101 \\
        &0111011100100011111100011101111\\ \hline 
$y_{7,1}$&011011011010001010001100010110011101000011100101011 \\
         &001111000011110110110000001011011100101110110011011\\
         &0001101011110101010011111011011\\ \hline 
$y_{7,2}$&100011011100010100100111010101000000100000101010011\\
        &111110101101110110010101100110000001111111001011110 \\
        &1111110101011000011101011010111\\ \hline 

$y_{8,1}$&111101011010111000011010101111110111101001111111000 \\
         &000110011010100110111011010111111100101010000010000\\
         &0010101011100100101000111011000\\ \hline 
$y_{8,2}$&011011011111001010101111010110001101100110111010011\\
        &101101000000110110111100001111001101010011100001011 \\
        &1001101000110001010001011011011\\ \hline 
$y_{9,1}$&011111111111111111101111111111111111110111111111111 \\
         &111111011111111111111111101111111111111111110111111\\
         &1111111111110111111111111111111\\ \hline 
$y_{9,2}$&100110111000010111010011011100001011101001101110000\\
        &101110100110111000010111010011011100001011101001101 \\
        &1100001011101001101110000101110\\ \hline 
\hline  
\end{tabular}
\end{table}





\end{document}